\documentclass [12pt]{article}
\usepackage{graphics}
\usepackage{epsfig}
\usepackage{array}
\usepackage{hangcaption}
\title{{\large\bf Nonlinear Supersymmetric Effective Lagrangian and
Goldstino Interactions at High Energies}}

\author{Taekoon Lee\footnote{email: tlee@physics.purdue.edu}$^{*(a,b)}$ and
Guo-Hong Wu\footnote{email: wu@physics.purdue.edu}$^{\dagger(a)}$
         \\
  {\it Department of Physics, Purdue University, West Lafayette, IN 47907
  }$^{(a)}$\\
  {\it Department of Physics, Seoul National University,
  Seoul, 151-742, Korea}$^{(b)}$\footnote{Present address.}
  }

\date{}
\begin{document}
\maketitle
\begin{abstract}
We show that 
the nonlinear supersymmetric effective lagrangian can
be used for model-independent {\it parameterization} of the 
light gravitino scattering amplitude at energies up to and
above  the soft supersymmetry-breaking masses. 
This provides the  most  convenient framework for 
systematic studies of goldstino phenomenology 
both at low energies and in high energy colliders.
\end{abstract}

\def\thepage{PURD-TH-98-10, SNUTP-98-135}
\thispagestyle{myheadings}
\newpage
\pagenumbering{arabic}
\addtocounter{page}{0}
\newcommand{\be}{\begin{equation}}
\newcommand{\ee}{\end{equation}}
\newcommand{\bear}{\begin{eqnarray}}
\newcommand{\eear}{\end{eqnarray}}

\section{Introduction}

The existence of a very light gravitino is one of the
characteristic features of low energy supersymmetry (SUSY)
breaking \cite{dine}.  If SUSY
breaking occurs in the order of the weak scale, the gravitino
becomes superlight with its mass several orders of magnitude smaller than
 an  electron Volt.
A superlight  gravitino could have important consequences in many  areas
such as cosmology, astrophysics,  and collider physics, and its
phenomenology has been  studied by many  authors \cite{many1,many2} since  
the early works of Fayet \cite{fayet}. 
One of the recent considerations on superlight gravitino
that is of particular interest to us is gravitino production in
high energy collisions.
As the typical energy of the process is well above 
the gravitino mass, one can replace the dominant, longitudinal 
component of the gravitino with the goldstino \cite{ET}.

In globally supersymmetric theories, two approaches exist for
obtaining the goldstino coupling to 
the fields of the minimal supersymmetric standard model (MSSM).
The  model-specific approach is to integrate out the high energy 
modes in a given SUSY breaking model
 down to a certain scale above the soft SUSY breaking masses in MSSM.
In general, the interactions of the goldstino with the MSSM fields are given in
nonderivative form, and are model-dependent, though the  vertices with a single
goldstino  are fixed by the Goldstino Goldberg-Treiman (GT) relation and
are model-independent. 
As shown in Ref.\cite{BFZ1}, the correct form of
goldstino-matter interaction is obtained only after careful treatment
of the Feynman diagrams that involve the nondecoupling of heavy particles.

The model-independent approach of the nonlinear effective 
lagrangian\cite{nll1,WB,CL,CLLW} is
based on a nonlinearly realized supersymmetry in the SUSY breaking sector,
 and provides the most general
interactions between the goldstino and a given set of fields consistent  with 
spontaneous SUSY breaking. Here the goldstino couples derivatively, and the
model-dependence on the  underlying  SUSY breaking mechanism appears as
undetermined coefficients of the derivative expansion of goldstino fields.
A clear advantage of this approach  is the decoupling of 
heavy particles from the goldstino emission processes.
The goldstino low energy theorem \cite{LET} follows automatically in the
nonlinear SUSY realization. 

 In our previous work \cite{CLLW}, we provided the rules for constructing
nonlinear SUSY invariant operators describing goldstino couplings to
matter and gauge fields in MSSM. The applications given there focus
on processes involving the goldstino and the standard model (SM)
 particles only.  
Such a nonlinear goldstino lagrangian for the SM
can also be obtained by explicitly integrating out the heavy SM superpartners
in a given model, as has recently been done by the authors of Ref.\cite{BFZ3} 
in their study of superlight gravitino production at $e^+e^-$ and hadron 
colliders when the SUSY particles are too heavy to be produced.
 It will be interesting and important to study also goldstino production
at energies near or above the sparticle masses.
 In the nonlinear SUSY lagrangian framework, however, such a study 
has not been done. 
It is our purpose here to extend our previous analysis to the high
energy regime.

  After a brief review of the nonlinear SUSY realization on the
MSSM fields, we study two types of scattering processes.
As will be shown, nonlinear SUSY invariance plays an important role
in constraining the structure of goldstino interactions.

\section{The nonlinear effective lagrangian}

 There exist in the literature several different 
approaches in obtaining the  nonlinear SUSY effective lagrangian 
\cite{nll1,CL,CLLW}. Here we take the
approach of Refs.\cite{CL,CLLW} which has an advantage over others
in that
the effective lagrangian can be written directly in terms of component
fields rather than  superfields as in other approaches.  
Because of this, it is straightforward to use this method to
find the most general SUSY invariant local  operators for a given set of fields.
One may refer to Ref.\cite{CL,CLLW} for details.

The nonlinear effective lagrangian is given in the form:
\begin{eqnarray}
\label{IEFF}
I_{\rm eff}=\int d^4x \,\, det A \,\, {\cal L}_{\rm eff}({\cal D}_\mu \chi,
{\cal D}_\mu \bar{\chi}, \phi^i, {\cal D}_\mu \phi^i, {\cal F}_{\mu\nu})
\end{eqnarray}
where ${\cal L_{\rm eff}}$ is a gauge invariant
function of the standard realization basic building blocks defined by
\bear
{\cal D}_\mu \chi &=& (A^{-1})_\mu~^\nu \partial_\nu\chi~~, \nonumber \\
{\cal{D}}_\mu\phi^i&=&(A^{-1})_\mu~^\nu (\partial_\nu \delta_{ij} +T^{a}_{ij}
A^{a}_{\nu}) \phi^j~~, \nonumber \\
{\cal F}_{\mu\nu}^a &=&
(A^{-1})_\mu~^\alpha (A^{-1})_\nu~^\beta F_{\alpha\beta}^a~~,
\eear
where $\chi$ is the goldstino, 
$\phi^i$ denotes generic scalar or fermion fields, 
$A^{a}_{\nu}$ and $F_{\alpha\beta}^a$ are respectively the gauge field 
and filed strength tensor, and  $T^{a}$ the gauge group generators.
The goldstino self-interaction is simply given by the
Volkov-Akulov action ${\cal{L}}_{AV}$  \cite{AV},
\be
{\cal{L}}_{AV}=-F^2~\det{A}
\ee
where  $F$ is the goldstino decay constant and
 $A$ is the Volkov-Akulov vierbein  defined by
\be
A_\mu^\nu = \delta_\mu^\nu + \frac{i}{2F^2}\chi
\stackrel{\leftrightarrow}{\partial}_\mu\sigma^\nu\bar{\chi} \;.
\ee

  It is convenient to catalog the terms in the effective Lagrangian,
${\cal L}_{\rm eff}$, by an expansion in the number of
goldstino fields which appear when the Volkov-Akulov vierbein is set to
 unity.  Then we have
\be
{\cal L}_{\rm eff}= \left[{\cal L}_{(0)}+ {\cal L}_{(1)}+
{\cal L}_{(2)}+\cdots \right] ~,
\ee
where the subscript $n$ on ${\cal L}_{(n)}$ denotes that each independent
SUSY invariant operator in
that set begins with $n$ Goldstino fields.

  For the MSSM fields,   ${\cal L}_{(0)}$  is obtained  from the MSSM 
lagrangian by replacing the
ordinary gauge covariant derivatives with the SUSY-gauge covariant
 derivatives and the ordinary field strength tensor with  the SUSY covariant
 field  strength tensor:
\be
{\cal L}_{(0)} = {\cal L}_{MSSM}(\phi,{\cal D}_\mu \phi, {\cal F}_
{\mu\nu}) .
\ee
Note that ${\cal L}_{(0)}$ is independent of the underlying SUSY breaking
dynamics, and  the goldstino dependence  arises only
from higher dimension terms in the matter covariant derivatives and
the SUSY covariant field strength tensor.
In particular, 
 the vertices having two on-shell goldstinos
arise from the SUSY and gauge invariant
kinetic terms for the matter and gauge fields,
\bear
{\cal L}_{(0)}^{\chi\chi} & = & 
- i \bar{\psi^i} {\bar{\sigma}}^{\mu} {\cal D}_{\mu} \psi^i
- ( {\cal D}^\mu \phi^i)^{\dag} ({\cal D}_{\mu} \phi^i)
- \frac{1}{4} {\cal F}_{\mu\nu}^a {\cal F}^{a\; \mu\nu}
\eear 

The terms in ${\cal L}_{(1)}$ describe, at the one goldstino
level, the standard single goldstino derivative coupling to 
the supercurrent,
\be
{\cal L}_{(1)} = \frac{1}{F} {\cal D}_{\mu}\chi^{\alpha}
J_{\alpha}^{\mu} + h.c.  
\ee
where $J_{\alpha}^{\mu}$ is the supercurrent.
Like ${\cal L}_{(0)}$,  ${\cal L}_{(1)}$ is model-independent.
Explicitly, the goldstino coupling to the fermion-scalar pair
$(\psi, \phi)$ and the gauge-gaugino pair $(A_{\mu}, \lambda)$
are given by
\be
  {\cal L}_{(1)} = \frac{1}{F} \partial_{\mu}\chi
 \sigma^{\nu}\bar{\sigma}^{\mu}\psi D_{\nu}\phi^{*}
 -\frac{1}{2\sqrt{2}F} \partial_{\mu}\chi
\sigma^{\nu}\bar{\sigma}^{\rho}\sigma^{\mu}\overline{\lambda} F_{\nu\rho}
 +\;\; h.c. + O(\chi^{3}) 
\label{e2}
\ee
where $D_{\mu} \phi$ is the gauge covariant derivative. 
Throughout this paper, the goldstinos are taken to be on-shell  external
particles.

   Equivalently, the single Goldstino interactions can be written in 
nonderivative form. For SUSY QED with a massless fermion carrying 
unit charge, this is simply given by
\be
{\cal L}_{(1)\mbox{{\scriptsize ND}}} =
\frac{m_{\phi}^{2}}{F} \chi \psi\phi^{*} +
\frac{im_{\lambda}}{\sqrt{2}F} \chi \sigma^{\mu\nu}\lambda F_{\mu\nu}
-\frac{ e m_{\lambda}}{\sqrt{2}F} \phi^{*}\phi\chi\lambda +
\;\;h.c. + O(\chi^{3})  \;\; ,
\label{nd}
\ee
where the first two are the standard GT trilinear
couplings, whereas the presence of
 the quartic term is required for the two forms of the single Goldstino
interactions to be equivalent \cite{LW}.

  Unlike ${\cal L}_{(0)}$ and ${\cal L}_{(1)}$, the rest of the
terms in the effective lagrangian are model-dependent.
The operators in ${\cal L}_{(2)}$, for example, can be written in
the form
\bear
{\cal L}_{(2)} &=& \frac{1}{F^2}{\cal D}_\mu \chi^\alpha {\cal
D}_\nu
\bar\chi^{\dot\alpha} M^{\mu\nu}_{\alpha\dot\alpha} + \cdots
\eear
where $M^{\mu\nu}_{\alpha\dot\alpha}$ denotes  operators composed of
the MSSM fields and containing arbitrary coefficients.

\section{Fermion-antifermion annihilation}

  In this section, we examine the validity of the nonlinear effective
lagrangian description of the fermion antifermion
annihilation process $\psi \bar{\psi} \rightarrow \chi \bar{\chi}$, both
in the zero energy limit and at energies above the sfermion mass.
For simplicity, the matter wyle fermion is assigned a conserved 
global $U(1)$ charge and is massless.  Nonlinear SUSY invariance will be 
seen to place very useful constraints on the low energy operators.

  We start with the model-independent sfermion exchange  contribution
shown in Fig.~1, using the nonderivative GT coupling of
Eq.(\ref{nd}). In the low energy limit, an expansion in $1/m^2_{\phi}$
gives rise to the effective local operators,
\be
   \frac{m^2_{\phi}}{F^2} (\chi \psi) (\bar{\chi} \bar{\psi})
   + \frac{1}{F^2} (\bar{\chi} \bar{\psi}) \partial^2 (\chi \psi)
   + \cdots \; ,
\label{sfex}
\ee
where terms suppressed by powers of $\partial^2/m^2_{\phi}$ have been
dropped. Note that 
nonlinear SUSY invariance forbids the presence of the first term 
which leads to incorrect energy dependence of the amplitude 
in the zero momentum limit of the goldstino.
The second term, on the other hand, is allowed by the goldstino
low energy theorem \cite{LET} and consistent with nonlinear SUSY
invariance as will be shown below. The low energy goldstino decoupling
as dictated by nonlinear SUSY realization thus 
requires the presence of new interactions to cancel out the first term above.

 As an example, we consider the toy model of Ref.\cite{BFZ2}.
The interaction terms of the model  relevant for our process
are given by
\bear \label{intf}
{\cal L}_{int} & = &  \frac{m^2_{\phi}}{F} (\chi \psi \phi^*
       + \bar{\chi} \bar{\psi} \phi )
       -\frac{m^2_{\phi}}{F^2} (\chi \psi) (\bar{\chi} \bar{\psi}) .
\eear
Note that the first two terms are the standard GT 
trilinear interactions which, after integrating out the sfermion,
give rise to the effective local operators of Eq.(\ref{sfex}).
The quartic operator in the model cancels the first term 
in Eq.(\ref{sfex}), leading to the effective low energy 
interaction well below the sfermion mass \cite{BFZ2},
\bear \label{Op1}
 {\cal O}_{ff} & = & \frac{1}{F^2} (\bar{\chi} \bar{\psi})
       \partial^2 (\chi \psi) .
\eear
This operator is consistent with nonlinear SUSY realization as we now
show.

\begin{figure} 
\begin{center}
\epsfig{file=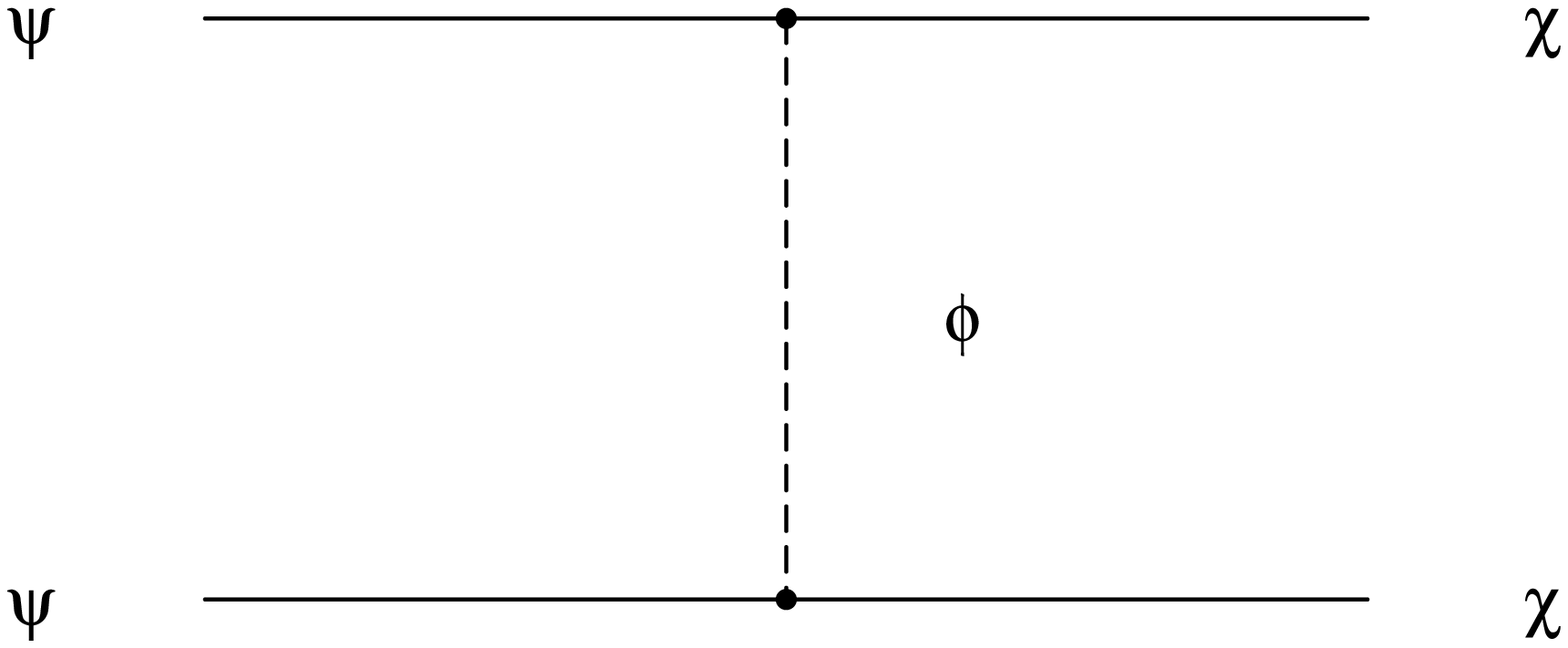, %
        height=2.3cm}
\end{center}
\isucaption{Goldstino pair production in fermion-antifermion annihilation
due to sfermion exchange.}
\end{figure}

  In the effective nonlinear goldstino lagrangian, two dimension-eight operators
can be written down to describe this low energy scattering.
They are given by \cite{CLLW,BFZ2}:
\begin{eqnarray} \label{NLfermion}
{\cal L}_{\psi\bar{\psi} \chi\bar{\chi} } &=& -\frac{1}{4F^2}
\left( \chi \stackrel{\leftrightarrow}{\partial}_\mu
\sigma^\nu \bar\chi\right) \left( \psi
\stackrel{\leftrightarrow}{\partial}_\nu \sigma^\mu \bar{\psi}  \right)
 +  \frac{C_{ff}}{F^2}\left( \psi \partial^\mu \chi\right)\left(\bar
{\psi} \partial_\mu \bar\chi\right) \; ,
\end{eqnarray}
where the first term comes from the SUSY invariant 
fermion kinetic term in ${\cal L}_{(0)}$
and the second is a model-dependent operator in ${\cal L}_{(2)}$ 
with an arbitrary coefficient. 
 The effective interaction of Eq.(\ref{Op1}) is recovered 
from Eq.(\ref{NLfermion}) when $C_{ff}=-2$ and is thus SUSY invariant.

 Now at energies of the  order of the sfermion mass, the scattering
amplitude develops a dependence on the mass of the superpartner.
 This is seen in the model of Ref.\cite{BFZ2} as arising from
the same two diagrams due to sfermion exchange and the contact term (Fig. 2).
Explicitly, the effective interaction is now given by,
\bear \label{Op2}
 {\cal M}_{ff}  & = & \frac{2}{F^2}
 (\partial_{\mu} \chi \partial^{\mu} \psi) (\bar{\chi} \bar{\psi})
   -  \frac{2}{F^2}  (\bar{\chi} \bar{\psi})
    \frac{\partial^2}{\partial^2 - m^2_{\phi}}
    (\partial_{\mu} \chi \partial^{\mu} \psi) ,
\eear
which reduces to Eq.(\ref{Op1}) in the zero energy limit.

 It is clear that the validity of the nonlinear effective lagrangian 
description of Eq.(\ref{NLfermion}) breaks down at the sfermion mass
scale, where the amplitude is expected to develop a dependence on the 
sfermion mass. 
 This seems to suggest that the nonlinear effective lagrangian is
a double expansion in $1/m_{\phi}$ and $1/F$, and that its usefulness
is limited below the relevant sparticle mass.

 Interestingly, however, this seemingly disappointing feature of the nonlinear 
effective lagrangian can be remedied by including the induced
contributions from  ${\cal L}_{(1)}$.
The contact term of Eq.(\ref{NLfermion}) plus the sfermion exchange 
diagram arising from ${\cal L}_{(1)}$ can be shown to give the same 
amplitude as in Eq.(\ref{Op2}) (see Fig. 2). 
Note that in the zero energy limit, the sfermion decouples from the 
nonlinear lagrangian and one is left with the contact interaction of
the first diagram on the R.H.S. of Fig. 2.
 The complete nonlinear effective lagrangian description of
${\cal L}_{(0)} + {\cal L}_{(1)} + {\cal L}_{(2)} + \cdots$ 
for goldstino emission
processes remains valid well above the soft SUSY breaking masses, 
and it is an expansion only in $1/F$.

\begin{figure}
\begin{center}
\epsfig{file=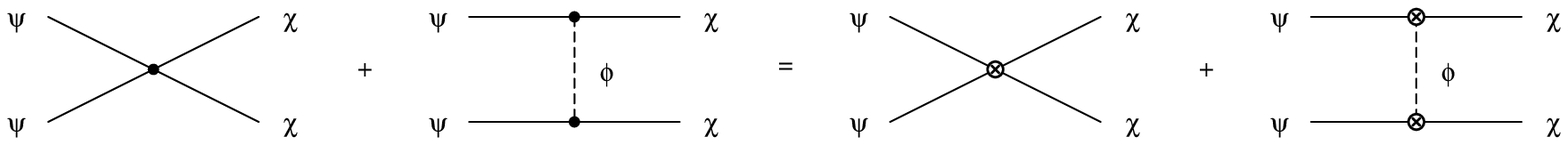,height=2.0cm,width=14cm}
\end{center}
\isucaption{Equivalence of amplitudes in fermion-fermion scattering.
            The dotted vertices are from the nonderivative lagrangian, while
            those on R.H.S. denoted by circled cross are from the nonlinear
            effective lagrangian.}
\end{figure}

\section{Photon-photon annihilation}

  As another example, we consider photon photon annihilation into 
a pair of goldstinos.
 Within the MSSM, this process receives contribution only from photino
exchange (Fig. 3).
 With the GT coupling of Eq.(\ref{nd}), one can easily find
the induced effective interaction at energies well below the photino 
mass $m_{\lambda}$,
\be \label{photino}
- \frac{i}{2F^2}  F^{\mu\nu} F_{\alpha\mu} \chi \sigma^{\alpha}
{\partial}_{\nu} \overline{\chi}
+ \frac{m_{\lambda}}{8F^2} \chi\chi (F_{\mu\nu}F^{\mu\nu}
  + \frac{i}{2} \epsilon^{\mu\nu\alpha\beta}F_{\mu\nu} F_{\alpha\beta} )
  + h.c.
\ee
where terms suppressed by powers of $E/m_{\lambda}$ are dropped.
The first term above corresponds to the dimension-eight operator
from the expansion of the photon kinetic term in the nonlinear effective
lagrangian. The second term can not be made invariant
under the nonlinear SUSY transformation, and has to be canceled 
by other contributions from outside of the MSSM.

  \begin{figure} 
  \begin{center}
  \epsfig{file=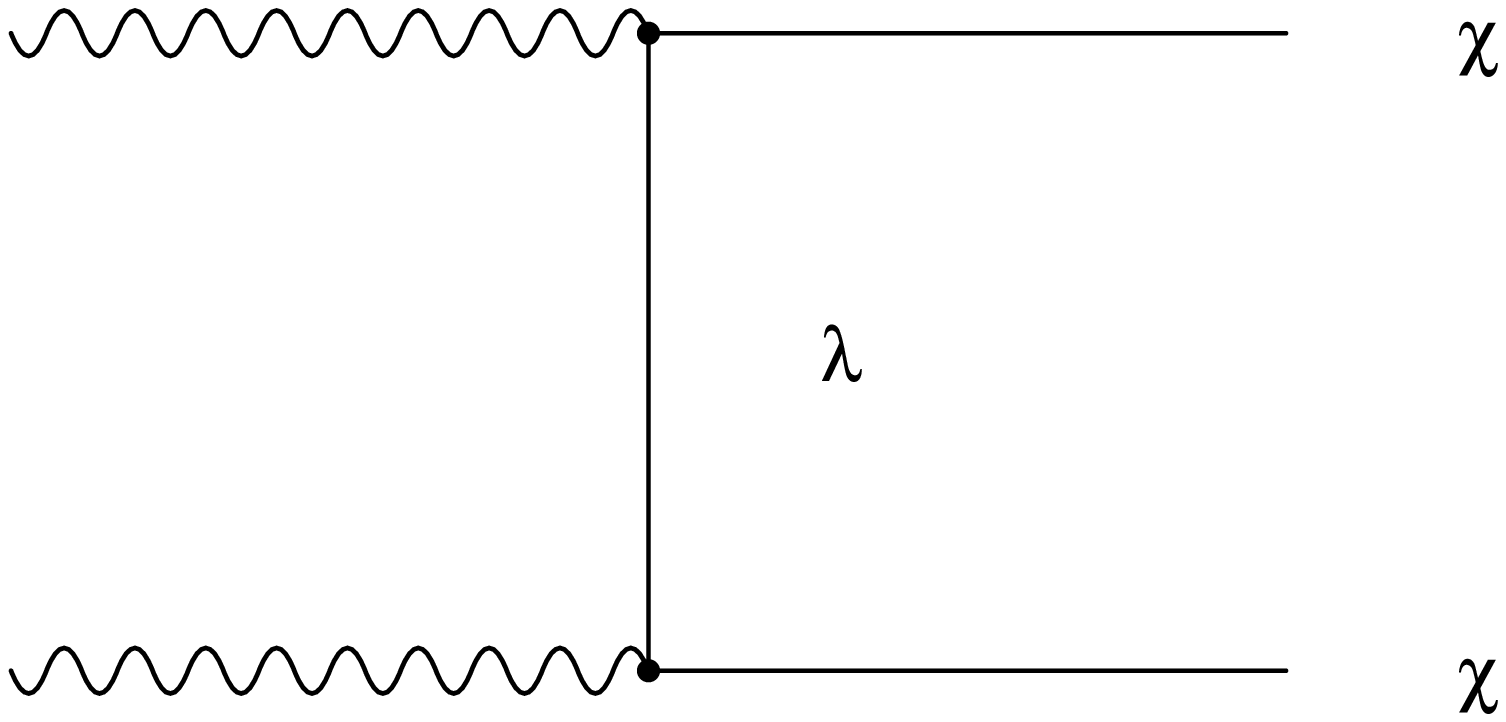,height=2.3cm}
  \end{center}
  \isucaption{Photino exchange in  photon-photon annihilation.}
  \end{figure}

  The low energy limit of this process has recently been studied in 
a toy model concerning the energy dependence of the cross section 
\cite{BFZ1}.
In this model, beside the standard GT coupling of Eq.(\ref{nd}),
there also exist  a scalar $S$  and a pseudoscalar $P$ coupled
to two goldstinos. The $S$ and $P$ are the scalar partners
of the goldstino. The relevant terms of the lagrangian in the model
are given by
\bear \label{SP}
{\cal L}_{int} & = &
 \frac{i}{\sqrt{2}} \frac{m_{\lambda}}{F} \chi
  \sigma^{\mu\nu} \lambda F_{\mu\nu}
-\frac{1}{2\sqrt{2}} \frac{m_S^2}{F} S \chi \chi
+ \frac{i}{2\sqrt{2}} \frac{m_P^2}{F} P \chi \chi
  \nonumber \\ &&
  -\frac{1}{2\sqrt{2}} \frac{m_{\lambda}}{F}
[ S F_{\mu\nu} F^{\mu\nu}
 - \frac{1}{2}  \epsilon^{\mu\nu\alpha\beta} P F_{\mu\nu} F_{\alpha\beta} ]
+ h.c.
\label{e17}
\eear
where $m_S$ and $m_P$ are the masses of the $S$ and $P$.
 In the zero energy limit, the contribution due to 
$S$ and $P$ exchange cancels the SUSY noninvariant term 
from the photino exchange in Eq. (\ref{photino}).  
The total low energy interaction is now given by, 
\bear \label{Ophoton1}
 {\cal O}_{\gamma\gamma} & = &
- \frac{i}{2F^2}  F^{\mu\nu} F_{\alpha\mu} \chi \sigma^{\alpha}
{\partial}_{\nu} \overline{\chi} \; ,
\label{e18}
\eear
which is independent of the masses of the photino and the $S$, $P$.

  From the nonlinear goldstino lagrangian standpoint, 
as a consequence of the derivative coupling nature of the goldstino,
the heavy sparticles always decouple from the low energy processes
and one only needs to consider the contributions of the
contact interactions.
 At the dimension-eight level, there exists only one such 
operator responsible for the low energy process
$\gamma\gamma \rightarrow \chi \bar{\chi}$.
This operator arises from the photon kinetic term in 
${\cal L}_{(0)}$ and is exactly given by Eq. (\ref{e18}),
 in accord with the explicit model calculation.

 We now turn to the high energy limit of the process and study
the validity of the nonlinear lagrangian.
In the model of Ref.\cite{BFZ1}, the contributing diagrams are
due to the exchange of photino, $S$, and $P$ respectively
(see Fig. 4).
Using the interactions of Eq. (\ref{e17}), the 
high energy amplitude is found to be 
\bear
{\cal M}_{\gamma\gamma}^{ND} & = &
- \frac{i}{2F^2}  \left( F^{\mu\nu} \chi \right ) \sigma^{\alpha}
     \frac{m^2_{\lambda}}{m^2_{\lambda}-\partial^2}
    \left[ ( {\partial}_{\nu} \overline{\chi}) F_{\alpha\mu} \right]
\nonumber \\
&& - \frac{m_{\lambda}}{8F^2} \left( F_{\mu\nu} \chi \right)
\frac{m^2_{\lambda}}{m^2_{\lambda}-\partial^2}
 \left[ \chi ( F^{\mu\nu} + \frac{i}{2} \epsilon^{\mu\nu\alpha\beta}
               F_{\alpha\beta} ) \right ]  + h.c. \nonumber \\
&& + \frac{m_{\lambda}}{8F^2} \left[
          F^{\mu\nu} F_{\mu\nu}  \frac{m_S^2}{m_S^2 - \partial^2}
          (\chi \chi)
    + \frac{i}{2} \epsilon^{\mu\nu\alpha\beta} F_{\mu\nu} F_{\alpha\beta}
            \frac{m_P^2}{m_P^2 - \partial^2}
          (\chi \chi) \right] + h.c. \; ,  
\label{e20}
\eear
 which reduces to Eq.(\ref{Ophoton1}) in the zero momentum limit.

  In the framework of the nonlinear goldstino lagrangian, the contact
contribution from the model-independent interaction in ${\cal L}_{(0)}$ 
(see Eq.(\ref{e18}) )
does not give the complete scattering amplitude at high energies.
Including the photino exchange contribution arising from 
${\cal L}_{(1)}$ now becomes necessary but is not sufficient by itself.
The toy model lagrangian of Eq.(\ref{SP}) suggests that 
the goldstino couplings to its scalar partners be included in the 
model-dependent piece of the nonlinear lagrangian ${\cal L}_{(2)}$,
which will give new contributions to the high energy photon-photon
annihilation process.
The coefficients of the derivative couplings of the goldstino to 
$S$ and $P$ can be determined in the model from the decay amplitudes
of $S \rightarrow \chi\chi$  and $P \rightarrow \chi\chi$.
They are given by  
\bear
{\cal L}_{(2)}^{S,P} & = & -  \frac{1}{\sqrt{2}F} (S -iP)
\partial_{\mu} \chi \partial^{\mu} \chi + h.c. \; .
\label{e21}
\eear

 The high energy scattering amplitude can now be calculated in the 
nonlinear lagrangian by including the contact contribution 
(Eq.~(\ref{e18})), the photino exchange, and the $S$ and $P$ 
exchanges.
It is given by   
\bear
{\cal M}_{\gamma\gamma}^{D}&= & - \frac{i}{2F^2}  F^{\mu\nu}
 F_{\alpha\mu} \chi \sigma^{\alpha}
{\partial}_{\nu} \overline{\chi}
\nonumber \\
&& - \frac{i}{2F^2}  \left( F^{\mu\nu} \chi \right ) \sigma^{\alpha}
     \frac{\partial^2}{m^2_{\lambda}-\partial^2}
    \left[ ( {\partial}_{\nu} \overline{\chi}) F_{\alpha\mu} \right]
\nonumber \\
&& - \frac{m_{\lambda}}{8F^2} \left( F_{\mu\nu} \chi \right)
\frac{\partial^2}{m^2_{\lambda}-\partial^2}
 \left[ \chi ( F^{\mu\nu} + \frac{i}{2} \epsilon^{\mu\nu\alpha\beta}
               F_{\alpha\beta} ) \right]  + h.c. \nonumber \\
&& + \frac{m_{\lambda}}{8F^2} \left[
          F^{\mu\nu} F_{\mu\nu}  \frac{\partial^2}{m_S^2 - \partial^2}
          (\chi \chi)
    + \frac{i}{2} \epsilon^{\mu\nu\alpha\beta} F_{\mu\nu} F_{\alpha\beta}
            \frac{\partial^2}{m_P^2 - \partial^2}
          (\chi \chi) \right] + h.c. \; .
\label{e22}
\eear
Though the two amplitudes in Eq.~(\ref{e20}) and Eq.~(\ref{e22}) 
come from completely different schemes, it is straightforward
 to see that they are indeed identical (see Fig. 4). 
 Note that when some of the superpartners (say $S$ and $P$)
become very heavy compared to the typical energy transfer in the process,
these particles automatically decouple in the nonlinear lagrangian and one
only needs to consider the contact interaction and the diagram
involving the light superpartner (say $\lambda$) exchange.
For the nonderivative goldstino coupling in the model of 
Ref.\cite{BFZ1}, the heavy particles do not decouple and 
need to be included in the diagrams.
We have thus explicitly shown that the nonlinear goldstino lagrangian 
provides a valid description of the annihilation process 
$\gamma \gamma \rightarrow \chi \chi$ at energies up to and
above the soft breaking sparticle masses. 
Contributions from ${\cal L}_{(0)}$, ${\cal L}_{(1)}$, and ${\cal L}_{(2)}$
all need to be taken into account at high energies.

\begin{figure}
\begin{center}
\epsfig{file=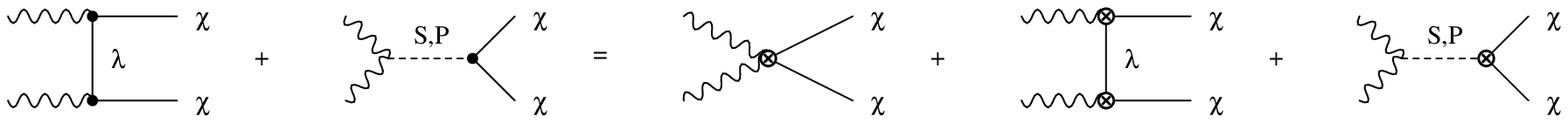,height=1.8cm,width=16cm}
\end{center} 
\isucaption{Equivalence of amplitudes in photon-photon annihilation.
           The dotted vertices are from the nonderivative lagrangian, while
            those on the R.H.S. denoted by circled cross are from the 
           nonlinear effective lagrangian.}
\end{figure}

\section{Conclusion}

If SUSY is broken around the weak scale, the accompanying gravitino
will be superlight, and direct gravitino production in
high energy collisions becomes feasible.
We have demonstrated that the nonlinear goldstino 
lagrangian provides a valid description of the goldstino interactions
both below and above the soft SUSY breaking masses, though the latter
requires the superpartners to be properly taken into account.
 At energies well below the soft masses, 
the superpartners decouple from the low energy theory and 
one recovers  the effective nonlinear goldstino lagrangian for the SM.
Generally speaking, particles much heavier than the typical energy of the 
goldstino always decouple in the nonlinear lagrangian approach, and 
one only needs to consider light particle effects. 
In contrast, with nonderivative goldstino coupling as is often the case
in explicit models of SUSY breaking, heavy particles do not decouple 
and their exchange effects need to be explicitly included in the diagrams.
The nonlinear goldstino lagrangian thus provides the most economical 
description of the supersymmetry breaking sector.

Before we understand the origin of supersymmetry breaking, 
the effective nonlinear goldstino lagrangian provides the most 
convenient framework for model-independent and systematic studies 
of goldstino
phenomenology at both low energies and in high energy collisions.
As all the operators in the nonlinear lagrangian are organized in expansions
of a single parameter, the inverse of the goldstino decay constant 
$1/F$, phenomenological studies of the goldstino can yield valuable 
information on the supersymmetry breaking scale, and may also provide helpful 
guidelines in seeking  the  realistic model of supersymmetry breaking.

\section{Acknowledgments}
 We thank Tom Clark and Sherwin Love for discussions, and Tao Han for
raising the question about the validity of the nonlinear lagrangian 
above soft SUSY breaking masses. GW is grateful for the hospitality
of the High Energy Theory Group at Brookhaven National Lab and of
the Fermilab Theoretical Physics Department.
 This work was supported in part by the U.S. Department of Energy
under grant DE-FG02-91ER40681 (Task B). One of the authors (T.L.) is
also supported in part by the KOSEF brain pool program.

\end{document}